\documentclass{iopjournal}
\usepackage[T1]{fontenc}
\usepackage{textcomp}
\usepackage{lmodern}
\usepackage{microtype}
\usepackage{graphicx}
\usepackage{placeins}
\usepackage{float}
\usepackage{array}
\usepackage{booktabs}
\usepackage{multirow}
\usepackage{url}
\hypersetup{colorlinks=true,allcolors=black}
\makeatletter
\renewcommand{\articletype}[1]{\vspace*{8mm}\noindent\reversemarginpar
\marginpar{\scriptsize \sf{\bfseries \MakeUppercase{#1}}}}
\makeatother

\begin{document}
\articletype{Paper}
\title{Leakage-Audited Benchmarking Reveals Limited Evidence for Cross-Subject Auditory-Evoked EEG Vowel Perception Decoding}
\author{Xiaoyang Li$^1$ and Zeyan Tao$^1$}
\affil{$^1$College of Medicine and Biological Information Engineering, Northeastern University, Shenyang, 110016, China}
\email{20246389@stu.neu.edu.cn (Xiaoyang Li); taozy@mails.neu.edu.cn (Zeyan Tao)}
\keywords{auditory EEG, vowel perception, cross-subject decoding, leave-one-subject-out validation, participant-level inference, reproducible benchmark}
\begin{abstract}
\textbf{Objective.} We evaluated cross-subject five-vowel perception decoding from auditory-evoked EEG with explicit trial definitions, participant-grouped validation, complete prediction coverage, and participant-level inference.

\textbf{Approach.} We parsed all Study 2 event tables from OpenNeuro ds006104 version 1.0.1 and analysed the consonant--vowel (CV) pair task. Marker--stimulus pairing converted 7,680 event rows into 3,840 independent CV trials. Excluding active-condition trials left 1,280 eligible control trials. A 400~\textmu{}V maximum across-channel peak-to-peak criterion rejected 186 epochs, retaining 1,094 epochs from 16 participants and 61 EEG channels. Thirteen unique implementations underwent leave-one-subject-out (LOSO) testing. Participant metrics came from 36,102 trial predictions across 33 complete replicas. Descriptive analyses assessed deep-model stability across five seeds, class-resolved errors, and retention sensitivity. An exploratory MDM analysis comprised 9,616 genuine model refits across training cohorts of 3--15 participants.

\textbf{Main results.} Two independent cohort-construction implementations preserved the same ordered 1,094 trials and labels (maximum tensor difference, $4\times10^{-12}$~V). Thus, corrected accounting changed the reported exclusion decomposition but not the analysed cohort or model inputs. Random Forest achieved the highest mean balanced accuracy: 21.474\% (95\% participant-bootstrap interval, 19.526--23.482\%; five-class chance, 20\%). The one-sided Wilcoxon, Bonferroni-adjusted, and sign-flip $p$ values were 0.090897, 1.000000, and 0.092791, respectively. All 13 implementations remained non-significant after multiplicity correction. Deep-architecture means were near chance, with substantial within-participant seed ranges and low trial-label agreement for several architectures. Across 3--15 training participants, refitted MDM means ranged from 20.553\% to 20.922\%, without a monotonic increase.

\textbf{Significance.} This dataset and protocol provide limited evidence for cross-subject five-vowel decoding. Explicit trial units, participant-grouped validation, complete seed reporting, prediction-level traceability, and multiplicity-adjusted inference sharpen the interpretation of small-cohort EEG benchmarks.
\end{abstract}

\section{Introduction}

Cross-subject EEG decoding tests whether a model trained on several participants generalizes to an unseen listener without participant-specific calibration. This demanding target is central to reusable neural-engineering systems. Auditory EEG combines low signal-to-noise ratio with participant-specific neurophysiology, acquisition variability, and few labelled trials. Analytical choices can therefore exert effects comparable to the expected decoding signal. Ambiguous trial definitions, participant leakage, held-out-data normalization, selective seed reporting, and multiple candidate pipelines can produce optimistic estimates of generalizable decoding \cite{herff2016,jayaram2018,defossez2023}.

Single-trial EEG classification depends critically on signal amplitude, trial count, feature dimension, and the unit of generalization. Event-related responses may be reliable after averaging yet weak at the individual-trial level \cite{blankertz2011}. Continuous-speech EEG studies have also identified phoneme-related neural structure \cite{diliberto2015}. Such encoding evidence addresses a different question from discrete cross-participant vowel classification. Filtering, rereferencing, artifact handling, epoch definition, and feature scaling shape both decoding performance and biological interpretation \cite{kessler2025}. Cross-subject evaluation therefore requires transparent preprocessing, training-fold estimation of all model-fitting and normalization quantities, and explicit disclosure of any unlabeled per-recording quality-control operation.

Validation uncertainty becomes prominent when the number of independent participants is small. Cross-validation estimates may have wide sampling error, while extensive selection on shared folds can favour an optimistic winner \cite{varoquaux2017,varoquaux2018,varma2006}. Circular feature selection or inspection of test behaviour compounds this risk \cite{kriegeskorte2009}. A persuasive benchmark integrates prespecified endpoints, participant-level uncertainty, multiplicity across implementations, technical-replica stability, and trial-level prediction consistency.

Public-data benchmarks also depend on an explicit mapping from event rows to experimental trials. OpenNeuro ds006104 contains several Study 2 tasks and TMS-target labels \cite{moreira2025,dataset2025}. The dataset report distinguishes single-phoneme, CV-pair, and word-level material, while each CV trial has both a TMS marker and stimulus row. Treating rows as trials, pooling tasks, or combining active and control conditions changes the analytical denominator before preprocessing. Reproducibility therefore requires the counting unit and exclusion rule at every transition from released events to model input.

BIDS and OpenNeuro provide structured metadata, versioning, and durable access for transparent reuse \cite{pernet2019,markiewicz2021}. Scientific trial definitions nonetheless remain analysis-specific. Traceability from retained epochs to source files and event rows distinguishes design exclusions from boundary, annotation, and signal-quality losses.

Traceability must continue from preprocessing to reported metrics. Exported predictions permit metric regeneration and reveal when several displayed labels share one implementation. Cross-subject benchmarks also require a clear hierarchy of replication. Held-out participants define inferential units, whereas random seeds are repeated realizations nested within an architecture. Collapsing these levels inflates apparent sample size and can conceal instability. Architecture, optimization, regularization, and initialization all influence deep EEG results \cite{lotte2018,roy2019}. We therefore mapped labels to unique implementations, retained every technical-seed replica, and derived participant metrics from trial predictions.

Near-chance performance can arise from uniformly weak, unstable, or class-biased decisions. Model means obscure trial-level seed agreement, vowel-specific errors, and sensitivity to participants with fewer retained trials. Training-size trends likewise require separate fitting within each cohort while preserving an untouched test participant. We therefore complemented the primary benchmark with seed-, class-, retention-, and training-cohort analyses.

Recent analyses of the same experiment pursued different targets. The original pilot emphasized TMS-related decoding effects \cite{comstock2024}. A recent independent preprint examined broader articulatory and sequence targets and highlighted confound vulnerability and weaker fine-grained decoding \cite{madishetty2026}. The present endpoint instead isolates five-vowel LOSO decoding in control-condition CV trials. Its scope is cross-subject vowel classification under this task and protocol.

Here, we evaluate five-way cross-subject vowel classification from stimulus-locked EEG in eligible Study 2 control-condition CV trials. We derive independent trials from all released Study 2 events and retain 1,094 epochs from 16 participants. Thirteen classical, covariance-based, and deep implementations are compared using participant-level LOSO testing and 36,102 trial predictions. We further quantify seed dependence, class-specific errors, retention sensitivity, and training-cohort effects across 9,616 MDM refits. The 13-implementation comparison defines the confirmatory family; the remaining analyses characterize uncertainty and boundary conditions.

\section{Methods}

\subsection{Source dataset, tasks, and event-row inventory}

This secondary analysis used OpenNeuro ds006104 version 1.0.1, a versioned BIDS release \cite{dataset2025,pernet2019,markiewicz2021}. Study 2 comprises 16 participants (S01--S16). We parsed 48 \texttt{events.tsv} files, with three task files per participant. The 21,776 rows comprised three tasks. \texttt{task-Words} contributed 7,386 rows (3,693 markers and 3,693 stimuli). \texttt{task-phonemes} contributed 7,680 rows (3,840 markers and 3,840 stimuli). \texttt{task-singlephoneme} contributed 6,710 rows (3,355 markers and 3,355 stimuli).

The benchmark focused on \texttt{task-phonemes}, which presents CV pairs. The separate \texttt{task-singlephoneme} files contained 220 trials for 14 participants, 110 for S01, and 165 for S13. These files comprised 1,830 consonant and 1,525 vowel stimuli. Task selection excluded all 3,355 single-phoneme trials.

\subsection{Marker-to-stimulus pairing and independent trials}

Rows within each selected \texttt{task-phonemes} file retained their released order. Each trial paired a \texttt{TMS} marker with the immediately following \texttt{stimulus} row. Valid pairs shared \texttt{tms\_target}, included a released trial number, and combined a consonant in \texttt{phoneme1} with an analysed vowel in \texttt{phoneme2}. The vowels were \texttt{a}, \texttt{e}, \texttt{i}, \texttt{o}, and \texttt{u}. All 3,840 pairs satisfied these criteria and were unique. Marker-to-stimulus latency was nominally 50~ms.

The stable identifier combined participant, task, and released trial number. Thus, 7,680 selected event rows defined 3,840 independent CV trials, with marker rows supplying paired metadata.

\subsection{Experimental-condition eligibility}

Eligibility followed the released BIDS \texttt{tms\_target} labels. The design excluded active-TMS trials labelled \texttt{lip} ($n=1{,}280$) or \texttt{tongue} ($n=1{,}280$). Trials labelled \texttt{control\_lip} ($n=640$) or \texttt{control\_tongue} ($n=640$) formed the analysis cohort. We retain the source designation ``control condition'' throughout.

Each participant contributed 240 selected CV trials: 160 active-condition trials and 80 eligible control-condition trials. Each vowel contributed 768 selected trials: 512 active-condition trials and 256 eligible control-condition trials. Condition eligibility was assigned before EEG preprocessing.

\begin{table}[H]
\caption{Complete global event-to-trial flow. The counting unit changes from event-table rows to independent CV trials only after one-to-one marker/stimulus pairing.}
\label{tab:flow}
\centering
\footnotesize
\begin{tabular}{@{}p{0.34\textwidth}rrp{0.21\textwidth}@{}}
\toprule
Stage & Excluded & Remaining & Counting unit \\
\midrule
All Study 2 event-table rows & 0 & 21,776 & event-table rows \\
Select \texttt{task-phonemes} rows & 14,096 & 7,680 & event-table rows \\
Pair marker/stimulus rows & 0 trials; 3,840 linked markers & 3,840 & independent CV trials \\
Select released control conditions & 2,560 & 1,280 & independent CV trials \\
Boundary/annotation check & 0 & 1,280 & eligible epochs \\
Peak-to-peak rejection ($>400$~\textmu{}V) & 186 & 1,094 & retained epochs \\
\bottomrule
\end{tabular}
\end{table}

\subsection{EEG preprocessing and epoch rejection}

MNE-Python loaded each participant's Study 2 \texttt{task-phonemes} EDF \cite{gramfort2014}. Each EDF exposes 62 signals: 61 EEG channels and one non-EEG \texttt{Status} trigger channel. The sidecar identifies CPz as the acquisition reference and AFz as ground. We retained the ordered 61 EEG channels and removed \texttt{Status}. Continuous data were resampled from 2,000 to 256~Hz, filtered from 0.5 to 40~Hz with a zero-phase FIR \texttt{firwin} filter, and rereferenced to the instantaneous average.

Eligible control stimuli were mapped to samples as \texttt{int(onset\_seconds $\times$ 256)}. Epochs spanned $-0.2$ to 1.0~s, with baseline correction over $-0.2$ to 0~s. MNE rounding produced 308 samples from $-0.19921875$ to 1.0~s. With annotation rejection enabled, every eligible epoch remained within EDF boundaries and clear of bad annotations.

For each eligible epoch, peak-to-peak amplitude was computed for every channel across all 308 samples. An epoch was rejected if the maximum across-channel peak-to-peak amplitude exceeded 400~\textmu{}V. Of 1,280 eligible epochs, 186 were rejected and 1,094 retained. Retained/rejected counts were: /a/, 222/34; /e/, 210/46; /i/, 228/28; /o/, 223/33; and /u/, 211/45. The trial-level preprocessing table records source file and row, released trial number, paired onsets, consonant, vowel, condition, maximum across-channel peak-to-peak amplitude, exclusion reason, and final status. Because each eligible stimulus trial generates one epoch, retained-trial and retained-epoch counts are numerically identical; we use ``retained epoch'' for preprocessing and ``retained test trial'' when describing prediction coverage.

\subsection{Cohort construction and numerical reproducibility}

Two independent implementations generated the analysis tensor from the same recordings and preprocessing specification. We harmonized participant labels (\texttt{S01} versus \texttt{sub-S01}) before comparing dimensions, ordered identifiers, labels, and every tensor element. Numerical agreement required a maximum absolute difference below $10^{-10}$~V. This comparison evaluated reproducibility of cohort construction and preprocessing.

\subsection{Feature representations and model definitions}

Classical models used 1,159 features per epoch: 305 log band-power values, 305 differential-entropy values, 183 Hjorth values, and 366 temporal statistics. The five frequency bands were 0.5--4, 4--8, 8--13, 13--30, and 30--40~Hz. Non-finite values were replaced by zero. After artifact rejection, label-free recording-level quality control computed channel variance across retained epochs and time. Channels with within-recording variance $z>3$ were zero-filled before classical feature extraction. This step used the unlabeled held-out recording and was distinct from inductive, fold-fitted normalization.

Covariance models used the retained 61-by-308 tensor without classical zero-fill. Each trial covariance was estimated with Ledoit--Wolf shrinkage and $10^{-8}I$ regularization \cite{ledoit2004}. The five covariance implementations were MDM, MDM-EA, TS-LDA, TS-SVM, and TS-SVM-EA \cite{barachant2013,congedo2017,he2020}. The three classical implementations were XGBoost, Random Forest, and LightGBM.

Deep models used a deterministic 16-block temporal representation of the 308 samples. Integer edges from \texttt{linspace(0, 308, 17)} were 0, 19, 38, 57, 77, 96, 115, 134, 154, 173, 192, 211, 231, 250, 269, 288, and 308. Each block was replaced by its mean. Five unique deep implementations were retained: EEGNet, CNN-1D, CNN-BiLSTM, CompactShallowNet, and EEG-Conformer \cite{lawhern2018,schirrmeister2017,song2023}. The EEG-Conformer projection used group normalization with four groups over 16 channels.

The additional label \texttt{EEGNet-FBCSP} instantiated the EEGNet class and produced identical fold metrics and trial predictions. It was therefore treated as an alias. The primary family comprised 13 unique implementations: three classical models, five covariance models, and five deep architectures.

\subsection{LOSO fitting and inductive normalization}

Each outer fold reserved one participant for final testing. Classical feature scaling estimated means and population standard deviations from the other 15 participants. The resulting fixed transform was applied to the held-out participant. This prespecified participant-grouped split prevented person-specific structure from crossing between training and test data \cite{varoquaux2017,varma2006}.

For MDM-EA and TS-SVM-EA, the Euclidean-alignment reference was the arithmetic mean of training-trial covariance matrices. After symmetrization and an eigenvalue floor of $10^{-10}$, its inverse square root transformed training and held-out covariances. The reference therefore derived exclusively from training-fold covariance statistics. MDM class prototypes were arithmetic means of training matrix logarithms; prediction minimized Frobenius distance in log-matrix space. Supplementary Methods provide the equations and pseudocode.

For each deep outer fold and seed $s$, \texttt{default\_rng}($s+1009h$) drew three of the remaining 15 participants without replacement for validation, where $h$ is the zero-based held-out-participant index. Sampling was unstratified, and each fold--seed split was shared across architectures. The other 12 participants formed the optimization set. Per-channel means and population SDs were estimated across optimization trials and 16 blocks. The fixed values standardized optimization, validation, and held-out arrays before clipping to $[-8,8]$. SDs below $10^{-6}$ were set to one. Training used seeds 42--46, Adam, learning rate $5\times10^{-4}$, and batch size 32. Inverse-frequency class-weighted cross-entropy used optimization labels; gradient norms were clipped at 5. Training allowed 150 epochs, with validation-loss early stopping at minimum improvement 0.1 and patience 20. Safe retries used learning rate $2.5\times10^{-4}$.

\subsection{Prediction aggregation and metric computation}

Participant metrics were computed from a trial-prediction table containing the trial identifier, held-out participant, true vowel, predicted vowel, model, and replica identifier. The three 500-tree classical models were fitted across all 16 LOSO folds from the same 1,159-feature matrix. Random state was 42. XGBoost used maximum depth 6 and learning rate 0.05; Random Forest used minimum leaf size 5; LightGBM used 31 leaves and learning rate 0.05. Training-fold scaling was estimated independently in every fold.

The five deterministic covariance-based implementations contributed one prediction replica each. Each of the five deep architectures contributed all five technical-seed replicas rather than a selected seed. The prediction table consequently contained 33 complete replicas: three classical, five covariance-based, and 25 architecture-by-seed deep replicas. Every replica covered the same 1,094 trial identifiers exactly once, yielding 36,102 prediction rows. Balanced accuracy and macro-F1 were computed first for each held-out participant and replica. For the primary architecture-level summaries, the five seed-specific participant metrics were then averaged with equal weight; non-deep implementations had one replica. Balanced accuracy was also recomputed directly from the prediction rows for all $33\times16=528$ participant--replica pairs and compared with the corresponding replica metrics.

All hyperparameters used in the R2 reconstruction were inherited from the archived R1 executable specification and were not retuned during the final benchmark. In the documented executable paths, outer-fold held-out-participant performance was not used for parameter or checkpoint selection. Records of the original parameter-choice rationale and prospective freeze date were unavailable.

\subsection{Primary endpoint and statistical analysis}

The primary endpoint was balanced accuracy, the unweighted mean of recall across five vowels; nominal chance was 20\%. Macro-F1 was descriptive. The held-out participant was the inferential unit ($N=16$). For each implementation, a one-sided Wilcoxon signed-rank test compared participant-level balanced accuracy with 20\%. Deviations within $10^{-10}$ percentage points of zero were set to zero. SciPy used \texttt{zero\_method=wilcox}, \texttt{method=auto}, and alternative \texttt{greater}. Bonferroni correction covered the 13 unique implementations.

For the numerically highest implementation, the sign-flip statistic was the equal-participant mean deviation in balanced accuracy from 20\%, evaluated using a one-sided Monte Carlo upper tail with 10,000 random sign flips (seed 42). The Monte Carlo $p$ value included the observed-statistic correction. Descriptive 95\% confidence intervals used 10,000 participant bootstrap resamples with PCG64 seeds beginning at 20260804. These intervals quantify uncertainty across participants; multiplicity-adjusted tests define primary inference.

\subsection{Exploratory training-cohort analysis}

We examined training-cohort size using deterministic MDM, enabling repeated fitting without optimization-seed variability. For each held-out participant, we sampled training subsets of 3, 5, 7, 9, 11, and 13 from the other 15 participants. At each non-full size, 100 distinct subsets were drawn with seed 20260804. The 15-participant condition used the complete training set. Class covariance prototypes were re-estimated for every subset before evaluating the untouched participant. This design yielded $16\times(6\times100+1)=9{,}616$ genuine model refits.

Balanced accuracy was calculated for every refit, averaged within held-out participant and training size, and summarized across 16 equally weighted participants. Pointwise confidence intervals used 10,000 participant bootstrap resamples with shared draws across training sizes. We also calculated participant-paired change from $n=3$ and within-participant SD across 100 training subsets. Variability is undefined at $n=15$, which has one complete training set. The 15-participant predictions matched the primary MDM predictions exactly. This single-model, unadjusted analysis was exploratory.

\subsection{Seed, error-pattern, and retention diagnostics}

Deep stability was summarized for every architecture-by-seed cell using its 16 held-out-participant metrics. Within-participant variability was quantified by the range across the five technical seeds. Prediction stability was evaluated separately as raw trial-label agreement for each of the ten unique seed pairs within an architecture. For seeds $s_1$ and $s_2$, agreement was defined over the same 1,094 retained trials as
\begin{equation}
A_{s_1,s_2}=\frac{1}{1{,}094}\sum_{i=1}^{1{,}094}\mathbf{1}\!\left\{\widehat y_i^{(s_1)}=\widehat y_i^{(s_2)}\right\}.
\end{equation}
Raw agreement was used as a descriptive measure of trial-level decision consistency and was not chance-corrected; its expected value depends on the two seeds' predicted-class marginal distributions. Seeds were technical replicas; participants remained the inferential units.

For the numerical-best primary implementation, a trial-pooled row-normalized confusion matrix was accompanied by vowel recall calculated separately for each participant; recall confidence intervals used 10,000 participant bootstraps. We also examined the two-sided Spearman association between retained test-trial count and participant balanced accuracy. A Theil--Sen line was included only as a descriptive visualization of this association.

Retention sensitivity summarized participants with at least 40 retained trials ($N=15$) or at least 50 ($N=13$), relative to all 16 participants. For each implementation and threshold, 10,000 paired participant-bootstrap resamples estimated uncertainty in the restricted-minus-full mean difference (seed 20260804). This post hoc diagnostic used the original out-of-subject predictions without model refitting. Class and retention analyses were descriptive and outside the primary hypothesis family.

\section{Results}

\subsection{Unit-aware cohort construction reveals participant-specific retention}

\begin{figure}[H]
\centering
\includegraphics[width=1.00\textwidth]{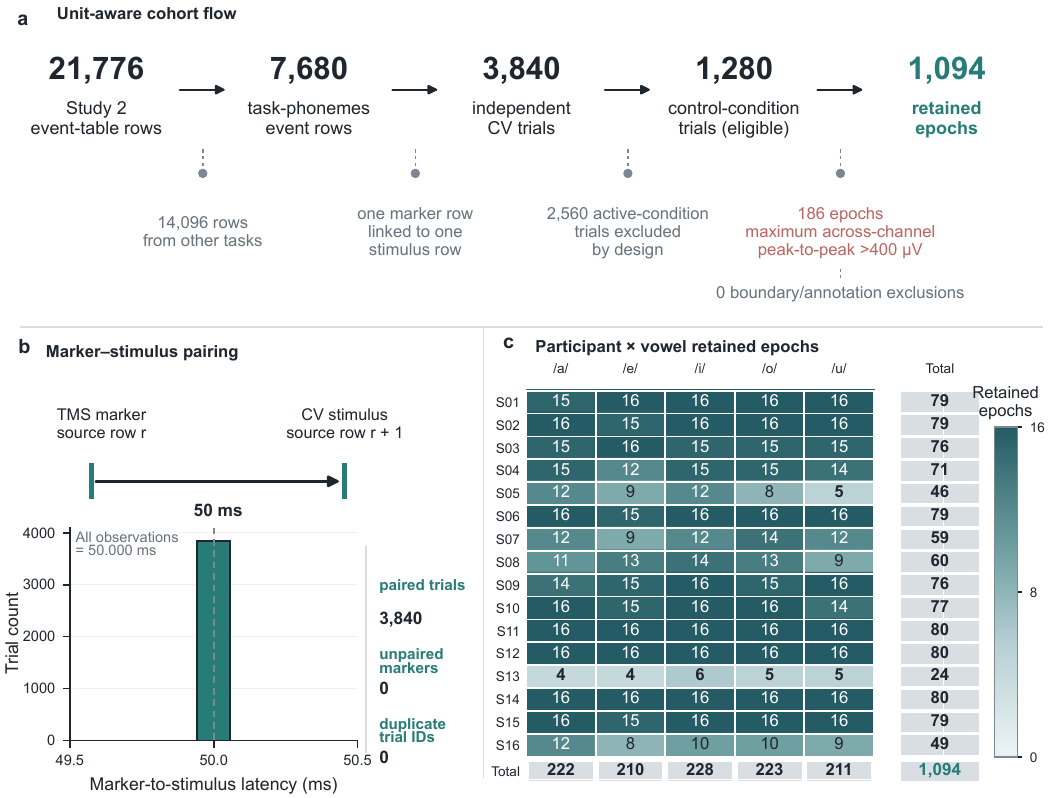}
\caption{\textbf{Unit-aware cohort construction and retained-epoch composition.} (a) Flow from 21,776 Study 2 event rows to 1,094 retained epochs. Numbers below arrows identify exclusions or linkage operations. The counting unit changes to independent CV trials after marker--stimulus pairing. Active-condition trials are design exclusions. No eligible epoch was excluded because of an EDF boundary or bad annotation (0 boundary exclusions and 0 annotation exclusions); 186 epochs were rejected when the maximum across-channel peak-to-peak amplitude exceeded 400~\textmu{}V. (b) Pairing structure and marker-to-stimulus latency for 3,840 independent CV trials. The dashed line marks 50~ms; every latency was 50.000~ms, and all markers paired uniquely. (c) Retained epochs by participant and vowel, with marginal totals. Each cell begins with 16 eligible control-condition trials; colour encodes 0--16 retained epochs. Source data accompany the figure.}
\label{fig:flow}
\end{figure}

We traced every released Study 2 row through task selection, trial pairing, condition eligibility, and EEG rejection (Figure~\ref{fig:flow}a). Of 21,776 event rows, 14,096 belonged to other tasks and 7,680 to \texttt{task-phonemes}. The selected rows comprised 3,840 markers and 3,840 stimuli, paired one-to-one into 3,840 independent CV trials. Every pair was adjacent, uniquely identified, and separated by 50.000~ms (Figure~\ref{fig:flow}b). The reduction from 7,680 rows to 3,840 trials therefore reflects the counting unit.

Condition selection excluded 2,560 active-condition trials by design and retained 1,280 eligible control-condition trials. No eligible epoch was lost because of an EDF boundary or bad annotation; 186 exceeded the 400~\textmu{}V maximum across-channel peak-to-peak criterion, leaving 1,094 retained epochs. Eligibility was balanced by construction, with 80 trials per participant and 256 trials per vowel, whereas post-rejection retention was heterogeneous (Figure~\ref{fig:flow}c). Participant totals ranged from 24 for S13 to 80 for S11, S12, and S14. Vowel totals occupied a narrower range, from 210 for /e/ to 228 for /i/, with /a/, /o/, and /u/ contributing 222, 223, and 211 epochs, respectively. The resulting analysis tensor had dimensions $(1{,}094,61,308)$. The two independent implementations preserved the same ordered trial identifiers and labels and differed by at most $4\times10^{-12}$~V. Accordingly, the corrected accounting changed the reported decomposition of exclusions but did not alter the analysed cohort, its labels, or any model input.

\subsection{Primary cross-subject performance remains close to chance}

\begin{figure}[H]
\centering
\includegraphics[width=1.00\textwidth]{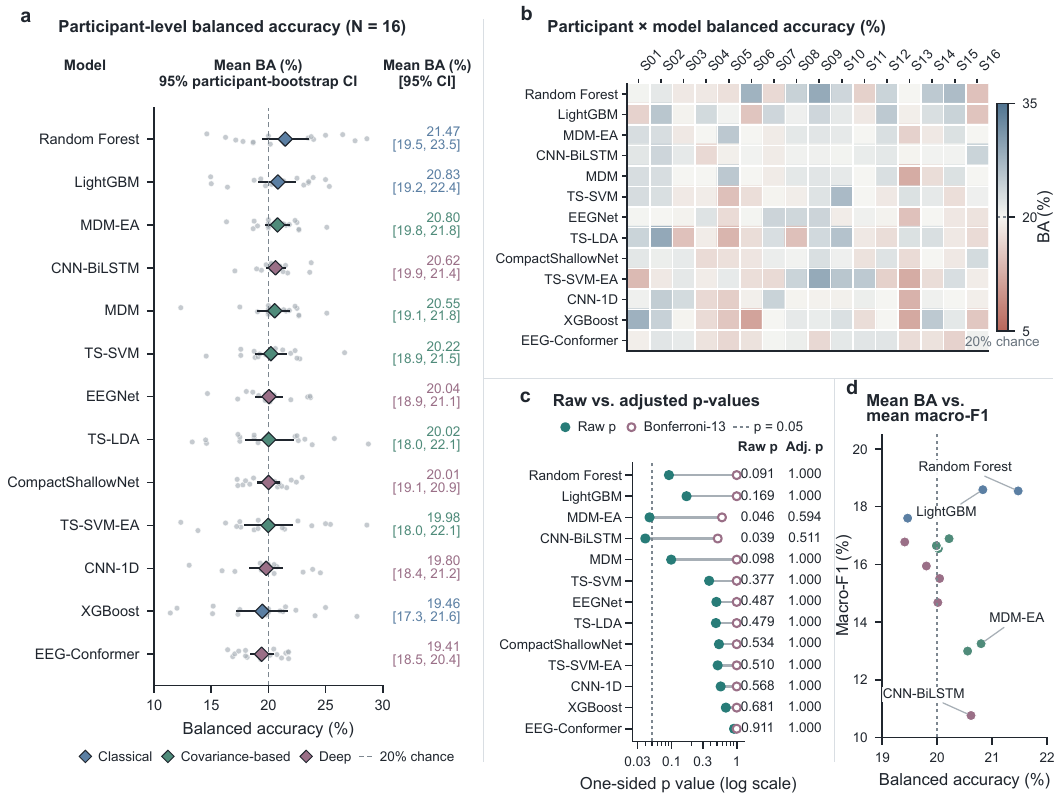}
\caption{\textbf{Primary participant-level cross-subject benchmark.} (a) Held-out-participant balanced accuracy for 13 unique implementations under LOSO validation. Grey points represent participants ($N=16$); diamonds show equal-participant means with 95\% intervals from 10,000 participant bootstrap resamples. Deep-architecture values are averaged across five technical seeds. Colour denotes model family; the dashed line marks five-class chance (20\%). (b) Participant-by-model balanced accuracy on a common colour scale. (c) One-sided Wilcoxon signed-rank $p$ values against 20\%, before and after Bonferroni correction across 13 implementations. The dashed line marks $p=0.05$. (d) Participant-mean balanced accuracy versus macro-F1, with selected models labelled to illustrate ranking differences. All adjusted tests remained above 0.05. Source data accompany the figure.}
\label{fig:primary}
\end{figure}

Participant-level balanced accuracy remained close to five-class chance across all 13 implementations (Figure~\ref{fig:primary}; Table~\ref{tab:models}). Means ranged from 19.409\% for EEG-Conformer to 21.474\% for Random Forest. Random Forest had the highest mean, with a 95\% participant-bootstrap interval of 19.526--23.482\%. Its one-sided Wilcoxon $p$ value was 0.090897, the Bonferroni-adjusted value was 1.000000, and the sign-flip $p$ value was 0.092791. MDM-EA and CNN-BiLSTM had nominal $p$ values of 0.045715 and 0.039340, with adjusted values of 0.594301 and 0.511425. All 13 adjusted tests remained above 0.05.

Participant-level variation exceeded the narrow separation between model means (Figure~\ref{fig:primary}b). Across 208 participant--model values, balanced accuracy ranged from 11.42\% to 28.75\%. Mean macro-F1 ranged from 10.76\% to 18.59\% and produced a different ordering (Figure~\ref{fig:primary}d). Random Forest combined the highest balanced accuracy with macro-F1 of 18.54\%. LightGBM had lower balanced accuracy (20.831\%) but the highest macro-F1 (18.586\%). Together, these metrics indicate weak multiclass performance with model rankings that depend on the endpoint.

\begin{table}[H]
\caption{Prediction-derived primary LOSO results for 13 unique implementations. BA values are participant-level mean and sample SD; macro-F1 is the mean of participant-level values. Deep participant metrics are averaged across five seed replicas. Bonferroni correction uses 13 tests.}
\label{tab:models}
\centering
\scriptsize
\begin{tabular}{llrrrrr}
\toprule
Implementation & Family & Mean BA (\%) & SD & Macro-F1 (\%) & Raw $p$ & Adjusted $p$ \\
\midrule
Random Forest & Classical & 21.474 & 4.226 & 18.544 & 0.090897 & 1.000000 \\
LightGBM & Classical & 20.831 & 3.326 & 18.586 & 0.169341 & 1.000000 \\
MDM-EA & Covariance-based & 20.799 & 2.114 & 13.250 & 0.045715 & 0.594301 \\
CNN-BiLSTM & Deep & 20.615 & 1.622 & 10.761 & 0.039340 & 0.511425 \\
MDM & Covariance-based & 20.554 & 2.835 & 12.993 & 0.098441 & 1.000000 \\
TS-SVM & Covariance-based & 20.216 & 2.762 & 16.889 & 0.377328 & 1.000000 \\
EEGNet & Deep & 20.044 & 2.379 & 15.510 & 0.487480 & 1.000000 \\
TS-LDA & Covariance-based & 20.022 & 4.327 & 16.544 & 0.479363 & 1.000000 \\
CompactShallowNet & Deep & 20.011 & 1.872 & 14.677 & 0.533953 & 1.000000 \\
TS-SVM-EA & Covariance-based & 19.984 & 4.376 & 16.646 & 0.510025 & 1.000000 \\
CNN-1D & Deep & 19.803 & 2.826 & 15.942 & 0.567661 & 1.000000 \\
XGBoost & Classical & 19.462 & 4.517 & 17.600 & 0.681117 & 1.000000 \\
EEG-Conformer & Deep & 19.409 & 1.964 & 16.774 & 0.910668 & 1.000000 \\
\bottomrule
\end{tabular}
\end{table}

\subsection{Implementation hierarchy and prediction coverage}

\begin{figure}[p]
\centering
\includegraphics[width=1.00\textwidth]{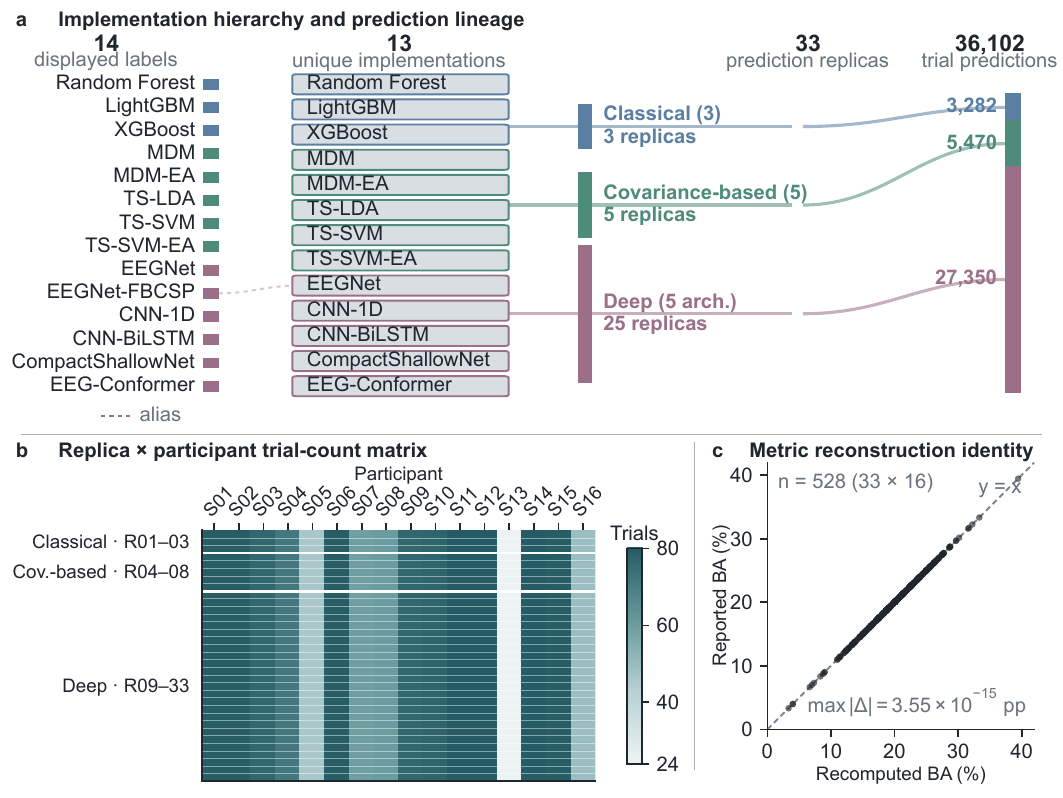}
\caption{\textbf{Implementation hierarchy and prediction coverage.} (a) Mapping of 14 displayed labels to 13 unique implementations, 33 prediction replicas, and 36,102 trial predictions. Colour denotes model family. Solid links show direct mappings; the dashed link identifies \texttt{EEGNet-FBCSP} as an EEGNet alias. Family totals comprise three classical replicas (3,282 predictions), five covariance-based replicas (5,470 predictions), and 25 deep architecture--seed replicas (27,350 predictions). (b) Retained-trial coverage by replica and held-out participant. Columns represent participants, rows represent replicas, and colour gives retained test trials; every row sums to 1,094. (c) Participant--replica balanced accuracy computed from predictions versus its summary value ($n=528=33\times16$). The dashed identity line is $y=x$; the maximum difference is $3.55\times10^{-15}$ percentage points. Source data accompany the figure.}
\label{fig:provenance}
\end{figure}

Fourteen displayed labels mapped to 13 unique implementations because \texttt{EEGNet-FBCSP} and EEGNet shared an architecture and outputs (Figure~\ref{fig:provenance}a). Three classical replicas contributed 3,282 predictions; five covariance-based replicas contributed 5,470; and 25 deep technical-seed replicas contributed 27,350. The resulting hierarchy comprised 33 complete replicas and 36,102 predictions, with EEGNet counted once.

Coverage was complete at both replica and participant levels. Every replica contained 1,094 retained trials. The $33\times16$ matrix reproduced participant totals of 24--80 trials in every row (Figure~\ref{fig:provenance}b). Balanced accuracy recomputed from predictions matched all 528 participant--replica summaries (Figure~\ref{fig:provenance}c). The maximum difference was $3.55\times10^{-15}$ percentage points. Model rankings, seed summaries, and inferential tests therefore derived from one complete prediction table.

\subsection{Deep-model averages conceal seed-dependent predictions}

\begin{figure}[H]
\centering
\includegraphics[width=1.00\textwidth]{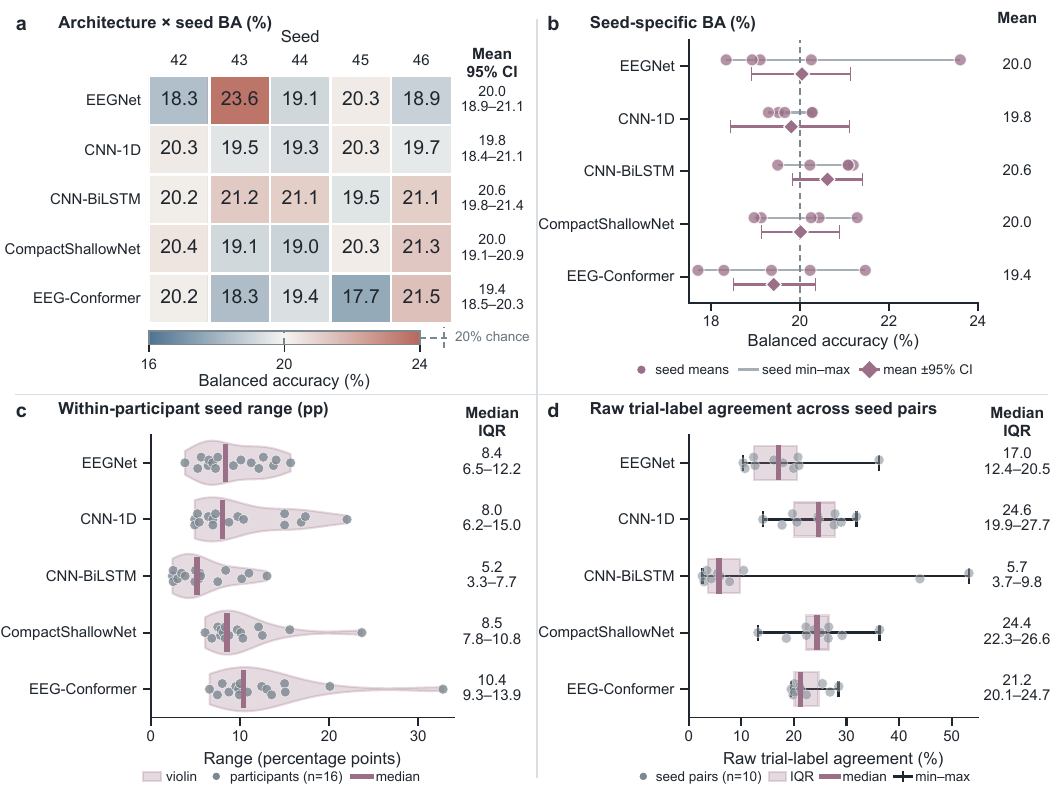}
\caption{\textbf{Deep-model performance and prediction stability across technical seeds.} (a) Participant-mean balanced accuracy for seeds 42--46 across five deep architectures. Cells show percentages; the right column gives equal-participant architecture means and 95\% participant-bootstrap intervals after within-participant seed averaging. The colour scale is centred at five-class chance (20\%). (b) Seed-specific architecture means (circles), minimum--maximum spans (grey lines), and architecture means with 95\% participant-bootstrap intervals (diamonds and coloured lines). Seeds are technical replicas; participants ($N=16$) are inferential units. (c) Within-participant balanced-accuracy ranges across five seeds, in percentage points. Points represent participants, violins show distributions, and coloured lines mark medians; right-margin labels report medians and interquartile ranges. (d) Raw, non-chance-corrected trial-label agreement percentages across ten seed pairs per architecture, using the same 1,094 trials. Points represent seed pairs, boxes show interquartile ranges, coloured lines mark medians, and whiskers span extrema. Source data accompany the figure.}
\label{fig:seed}
\end{figure}

We resolved all five deep architectures by technical seed (Figure~\ref{fig:seed}). Five architectures, five seeds, and 16 LOSO folds yielded 400 completed training units. Equal-participant architecture means remained close to chance: 20.044\% for EEGNet, 19.803\% for CNN-1D, 20.615\% for CNN-BiLSTM, 20.011\% for CompactShallowNet, and 19.409\% for EEG-Conformer. Every participant-bootstrap interval included 20\%. Seed-specific means spanned 18.340--23.603\% for EEGNet, 17.706--21.467\% for EEG-Conformer, and 19.498--21.195\% for CNN-BiLSTM (Figure~\ref{fig:seed}a,b).

Participant-resolved summaries revealed greater instability than architecture means (Figure~\ref{fig:seed}c). Median within-participant ranges were 8.4 percentage points for EEGNet, 8.0 for CNN-1D, 5.2 for CNN-BiLSTM, 8.5 for CompactShallowNet, and 10.4 for EEG-Conformer. Trial-label agreement captured complementary instability among similar aggregate scores. Across ten seed pairs and 1,094 trials per pair, median raw agreement was 17.0\%, 24.6\%, 5.7\%, 24.4\%, and 21.2\%, respectively (Figure~\ref{fig:seed}d). CNN-BiLSTM combined the narrowest performance range with the lowest prediction agreement. Aggregate stability therefore did not imply stable trial-level decisions.

\clearpage

\subsection{Increasing the training cohort does not produce a monotonic gain}

\begin{figure}[H]
\centering
\includegraphics[width=1.00\textwidth]{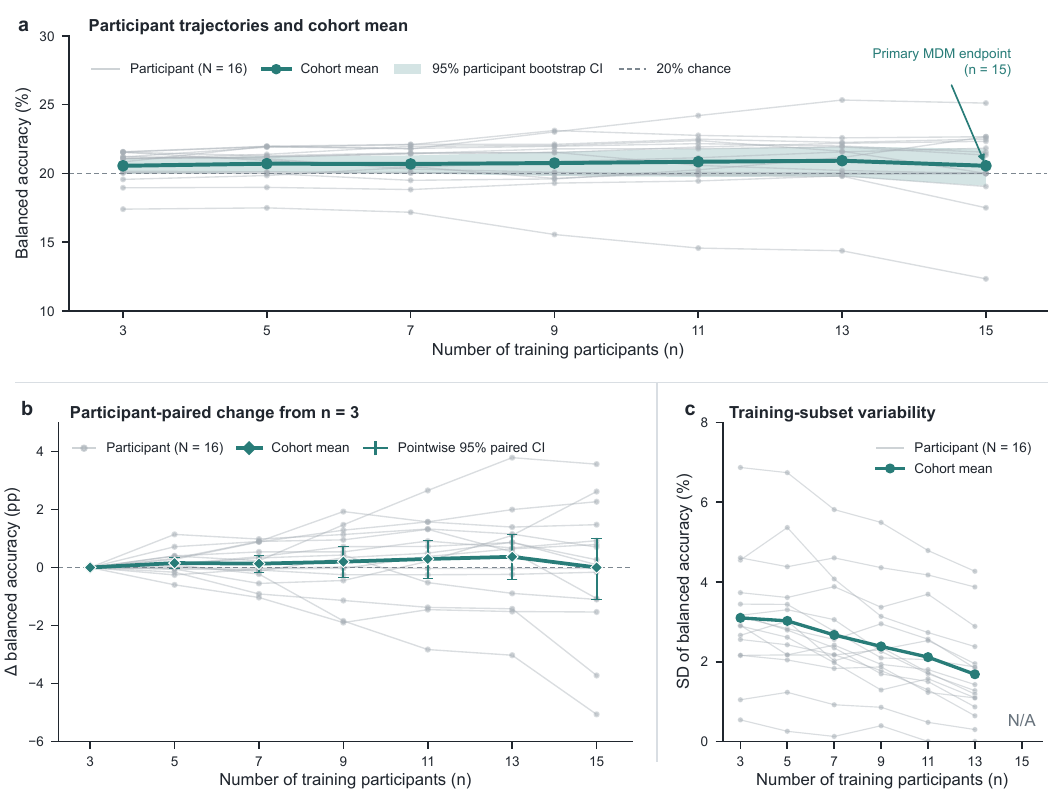}
\caption{\textbf{Exploratory MDM training-cohort analysis.} (a) Held-out-participant trajectories and equal-participant mean balanced accuracy for training cohorts of 3, 5, 7, 9, 11, 13, and 15 participants. Grey lines represent 16 participants; the teal line and shading show the mean and 95\% interval from 10,000 participant bootstrap resamples. The dashed line marks five-class chance (20\%). Values below $n=15$ average 100 unique subset refits; the labelled $n=15$ value is the primary MDM endpoint. (b) Participant-paired change from $n=3$. Teal diamonds and error bars show equal-participant means and pointwise 95\% paired-bootstrap intervals using shared draws across training sizes. (c) Within-participant SD across 100 sampled training subsets. Grey lines represent participants and the teal line their mean. The $n=15$ condition has one complete training subset. The exploratory analysis comprises 9,616 genuine model refits. Source data accompany the figure.}
\label{fig:learning}
\end{figure}

We fitted MDM separately across increasing training-cohort sizes while preserving an untouched participant (Figure~\ref{fig:learning}). Mean balanced accuracy was 20.553\%, 20.707\%, 20.686\%, 20.753\%, 20.848\%, 20.922\%, and 20.554\% for 3--15 training participants (Figure~\ref{fig:learning}a). The highest mean occurred at $n=13$; the $n=15$ endpoint returned to the primary MDM value. Relative to $n=3$, paired changes were 0.369 percentage points at $n=13$ and 0.001 at $n=15$ (Figure~\ref{fig:learning}b). Every pointwise 95\% paired-bootstrap interval included zero.

Larger training cohorts reduced sensitivity to subset composition (Figure~\ref{fig:learning}c). Mean within-participant SD across 100 subsets declined from 3.100 percentage points at $n=3$ to 1.684 at $n=13$. The single $n=15$ training set has no subset variability. Across 9,616 genuine model refits, full-cohort predictions matched the primary MDM predictions exactly. Reduced subset sensitivity therefore coexisted with no stable increase in mean balanced accuracy.

\subsection{Error pattern and retention sensitivity}

\begin{figure}[H]
\centering
\includegraphics[width=1.00\textwidth]{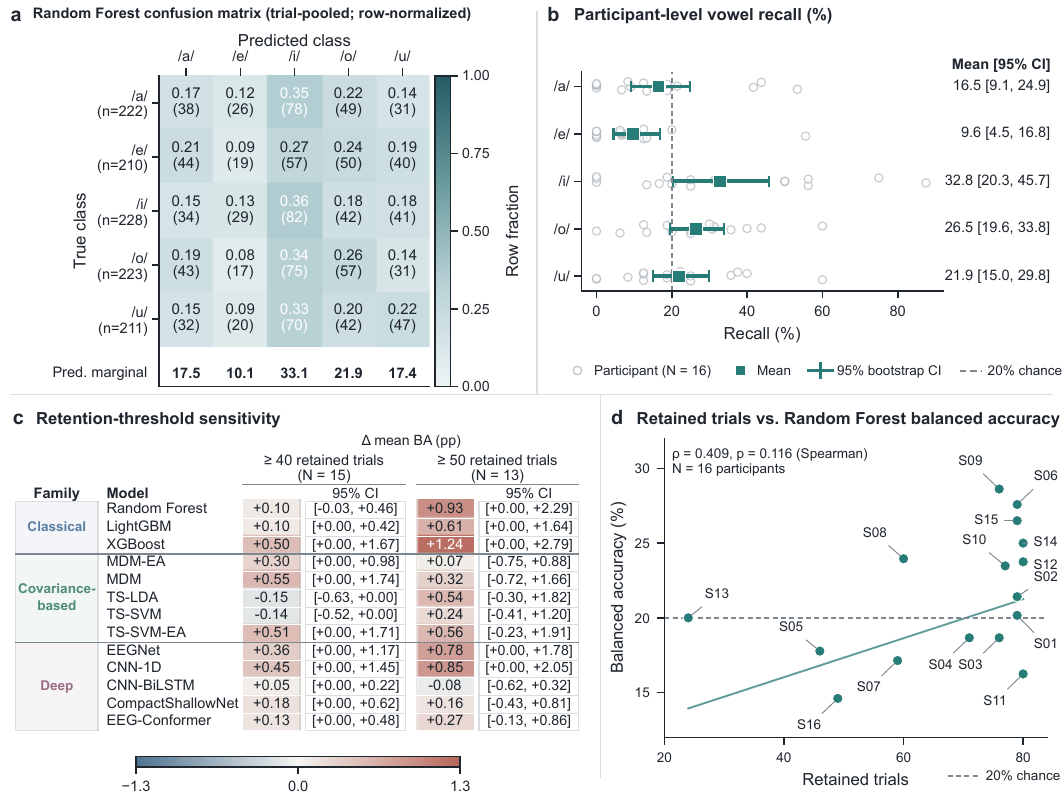}
\caption{\textbf{Class-resolved Random Forest errors and sensitivity to participant retention.} (a) Trial-pooled, row-normalized confusion matrix for the numerically highest implementation. Rows are true vowels and columns are predictions. Cells show row fractions and counts; margins give true-class totals and predicted-class percentages. (b) Vowel recall by held-out participant ($N=16$). Open circles represent participants, squares show equal-participant means, and horizontal intervals are 95\% participant-bootstrap intervals. The dashed line marks 20\%. (c) Change in model mean balanced accuracy among participants with at least 40 retained trials ($N=15$) or 50 trials ($N=13$), relative to all 16. Cells show changes and paired-bootstrap intervals from the original fitted models. (d) Retained trials versus Random Forest participant balanced accuracy. The solid line is a descriptive Theil--Sen fit over the observed range; the dashed line marks 20\%. The exploratory Spearman association was $\rho=0.409$ ($p=0.116$, two-sided). Panels are descriptive diagnostics. Source data accompany the figure.}
\label{fig:diagnostics}
\end{figure}

Random Forest errors were diffuse rather than diagonally dominant (Figure~\ref{fig:diagnostics}a,b). Class-conditional correct fractions were 0.17 for /a/, 0.09 for /e/, 0.36 for /i/, 0.26 for /o/, and 0.22 for /u/. The model assigned 362 of 1,094 trials (33.1\%) to /i/, which comprised 228 retained trials (20.8\%). Participant-level recall was 16.5\% [9.1, 24.9] for /a/, 9.6\% [4.5, 16.8] for /e/, and 32.8\% [20.3, 45.7] for /i/. Corresponding values were 26.5\% [19.6, 33.8] for /o/ and 21.9\% [15.0, 29.8] for /u/. The modest aggregate elevation was therefore class-dependent.

Retention-restricted summaries changed model means modestly (Figure~\ref{fig:diagnostics}c,d). Among 15 participants with at least 40 retained trials, changes ranged from $-0.15$ to $+0.55$ percentage points. Among 13 participants with at least 50 trials, changes ranged from $-0.08$ to $+1.24$ points. Random Forest changed by $+0.10$ and $+0.93$ points, with paired-bootstrap intervals of $[-0.03,+0.46]$ and $[+0.00,+2.29]$. These shifts preserved the multiplicity-controlled primary conclusion. Retained-trial count had a positive exploratory association with Random Forest balanced accuracy (Spearman $\rho=0.409$, $p=0.116$, $N=16$). These post hoc analyses characterize sensitivity to the evaluated participant set.

\section{Discussion}

This control-condition CV dataset provides limited evidence for cross-subject five-vowel decoding under the evaluated LOSO protocol. Random Forest exceeded nominal chance by 1.474 percentage points, with an interval spanning 20\% and non-significant Wilcoxon and sign-flip tests. MDM-EA and CNN-BiLSTM had nominal one-sided $p$ values below 0.05, but both remained non-significant after correction. Confidence intervals, participant-level tests, and multiplicity adjustment collectively outweigh the narrow numerical ordering among models.

A supplementary participant-level sensor-space analysis showed substantially greater participant-associated than vowel-associated structure in the retained EEG representation, providing descriptive context for the difficulty of cross-participant transfer. This analysis was not used for preprocessing selection, model selection, or confirmatory inference.

Unit-aware cohort reconstruction establishes the denominator for interpretation. Each CV trial has two event rows, and the dataset separates tasks and active/control conditions. The sequence was 7,680 selected rows, 3,840 paired trials, 2,560 design exclusions, 1,280 eligible trials, 186 amplitude rejections, and 1,094 retained epochs. Separating design exclusions from signal-quality loss revealed primarily participant-specific retention. Eligible counts were balanced, yet participant totals spanned 24--80 epochs while vowel totals remained between 210 and 228.

Complete prediction coverage further anchors model comparison. Fourteen displayed labels represented 13 implementations because EEGNet-FBCSP was prediction-identical to EEGNet. Each of 33 replicas covered the same 1,094 trials. All 528 participant--replica balanced accuracies matched values computed from 36,102 prediction rows to floating-point precision. This hierarchy preserves participants as inferential units and seeds as technical replicas nested within architectures.

Class-resolved results qualify the numerical Random Forest lead. Mean macro-F1 remained below 20\%, one-third of predictions were /i/, and recall ranged from 9.6\% for /e/ to 32.8\% for /i/. Balanced accuracy weights each true class equally, but cannot reveal whether errors are diffuse or concentrated. Combined with macro-F1 and confusion structure, it indicates a weak, class-dependent decision rule rather than consistent vowel discrimination.

Deep learning offered no stable alternative. Five-seed means placed every architecture near chance, while seed-specific means differed by up to 5.26 percentage points. Median within-participant ranges were 5.2--10.4 points. Median raw trial-label agreement across seed pairs was 5.7--24.6\%, despite evaluation on the same 1,094 trials. CNN-BiLSTM combined the smallest performance range with the lowest prediction agreement. Similar aggregate accuracy can therefore conceal markedly different trial-level decisions. Raw agreement should not be interpreted as an accuracy measure or as a formal comparison with a fixed 20\% null value. Complete seed reporting and prediction agreement are both important when individual decisions matter.

This instability limits the resolution of architecture comparisons. EEG classifiers offer distinct spatial, temporal, and covariance-sensitive inductive biases, yet conclusions depend on preprocessing, initialization, and participant sampling \cite{lotte2018,roy2019,kessler2025}. Here, between-model differences were smaller than many within-participant seed ranges, and the leading architecture depended on the metric. Participant distributions, confidence intervals, complete seeds, and trial-level agreement therefore convey more information than a rank-ordered leaderboard. Deep results characterize the inherited training and early-stopping specification and should not be interpreted as an exhaustive optimization of each architecture.

The statistical design reduces model-selection uncertainty while retaining the limits of a 16-participant cohort. Participant-grouped LOSO prevents trial leakage across people, but model ranks remain imprecise \cite{varoquaux2017,varoquaux2018}. Treating aliases as one implementation preserves the model family, and Bonferroni correction bounds its 13 confirmatory comparisons. These safeguards yield a single-dataset, leakage-audited estimate with participant-level uncertainty. External performance across sites or acquisition protocols remains untested \cite{varma2006,kriegeskorte2009}.

The exploratory training-cohort analysis showed no monotonic improvement across 3--15 participants. Among 9,616 genuine model refits, the largest increase from $n=3$ was 0.369 percentage points. Every paired interval included zero, and the $n=15$ endpoint returned to 20.554\%. Larger cohorts reduced subset variability, with mean within-participant SD declining from 3.100 to 1.684 points between $n=3$ and $n=13$. Thus, reduced subset sensitivity did not translate into higher mean discrimination over the evaluated range.

Retention heterogeneity contributed uncertainty without reversing the primary result. Excluding the lowest-retention participant changed Random Forest by 0.10 percentage points. Restricting analysis to 13 participants with at least 50 trials changed it by 0.93 points. Across models, shifts were small, directionally inconsistent, and preserved the multiplicity-controlled conclusion. The positive retention--performance correlation was non-significant. These post hoc analyses complement the prespecified all-participant endpoint.

The present target differs from prior analyses of this experiment. The pilot evaluated TMS-related decoding effects \cite{comstock2024}, whereas CIPHER examined broader articulatory and sequence targets \cite{madishetty2026}. Here, active-TMS trials were excluded, vowel identity was the target, and participant-level five-class balanced accuracy under LOSO was the endpoint. Differences in conditions, labels, validation units, and metrics preclude direct numerical comparison. The present result therefore addresses one specific cross-subject decoding question.

The study has six principal limitations. First, $N=16$ limits cross-subject precision, and retention is uneven, particularly for S13. Individual participants can materially influence model means in small cohorts \cite{varoquaux2018}. Second, the 400~\textmu{}V threshold is permissive. Trial-level rejection and retention sensitivity quantify its consequences, while threshold optimality remains unresolved. Different preregistered preprocessing choices may yield a different numerical benchmark \cite{kessler2025}.

Third, released events lack response time, response key, correctness, and button mapping. The $-0.2$ to 1.0~s window may therefore contain acoustic, decisional, or motor-related contributions. Fourth, classical channel-variance quality control uses unlabeled statistics from each recording, including the held-out recording. A deployment-oriented benchmark should predefine sensor handling independently of incoming-recording statistics.

Fifth, training-cohort analysis covers one deterministic covariance-based model. Seed, class, retention, supplementary sensor-space, and training-size analyses are descriptive and outside the multiplicity-adjusted family. Sixth, no independent dataset reproduces the CV-control protocol, leaving dataset-level generalization untested. BIDS organization and OpenNeuro versioning enable transparent reuse \cite{pernet2019,markiewicz2021}; protocol-matched external validation remains necessary.

A decisive next study should prespecify trial units, conditions, epochs, artifact rules, sensor policy, transformations, candidate models, and the primary metric. A larger, multisite sample would support site-aware validation and narrower uncertainty. An untouched protocol-matched cohort would separate development from final evaluation. Releasing response and button metadata would permit explicit analysis of perceptual and motor contributions. Candidate pipelines should export predictions and training outcomes, with every prespecified deep-model seed retained. These features align with recommendations for transparent neuroimaging analysis and limit circular evaluation \cite{nichols2017,kriegeskorte2009,varma2006}.

Within this dataset and protocol, near-chance means, class-dependent errors, seed-sensitive predictions, and non-significant inference provide limited evidence for reliable cross-subject five-vowel decoding. Explicit trial definitions, complete prediction coverage, participant-level inference, and separated confirmatory and descriptive analyses clarify weak EEG decoding signals. The evidence applies to control-condition CV trials and does not address imagined speech, overt speech, continuous listening, active-TMS effects, or online communication BCI.

\FloatBarrier
\ack{The authors thank the OpenNeuro dataset contributors and maintainers for making the source data publicly available.}

\section*{Ethics statement}
This secondary analysis used the publicly available OpenNeuro ds006104 dataset. The original documentation reports informed consent and approval by the UCLA Institutional Review Board (IRB\#21-000333). The study involved no new human-data collection.

\funding{This work received no external funding.}

\section*{Conflict of interest}
The authors declare no competing interests.

\roles{Xiaoyang Li performed the analysis and prepared the manuscript and reproducibility materials. Zeyan Tao contributed to trial-data curation, preprocessing verification, figure generation, and manuscript quality control. Both authors reviewed and approved the manuscript.}

\data{Source EEG and event data are available in OpenNeuro ds006104 version 1.0.1 (doi: 10.18112/openneuro.ds006104.v1.0.1). The public analysis archive is available at the record-specific doi: 10.5281/zenodo.21944149 (concept doi: 10.5281/zenodo.21805982; accessed 18 August 2026). It contains trial-level preprocessing data, retained-epoch identities, 36,102 primary predictions, and replica- and participant-level metrics. The archive also provides 9,616 training-cohort refits, waveform, distance and PCA data, figure source data, code, tests, and the software environment specification.}

\suppdata{Supplementary Methods provide trial-flow tables, channel and tensor definitions, normalization equations, prediction coverage, deep-seed results, training-cohort analyses, class and retention summaries, and descriptive sensor-space analyses.}


\begin{thebibliography}{99}
\bibitem{herff2016} Herff C and Schultz T 2016 Automatic speech recognition from neural signals: a focused review \textit{Front. Neurosci.} \textbf{10} 429. doi: 10.3389/fnins.2016.00429.
\bibitem{jayaram2018} Jayaram V and Barachant A 2018 MOABB: trustworthy algorithm benchmarking for BCIs \textit{J. Neural Eng.} \textbf{15} 066011. doi: 10.1088/1741-2552/aadea0.
\bibitem{defossez2023} Defossez A, Caucheteux C, Rapin J, Kabeli O and King J-R 2023 Decoding speech perception from non-invasive brain recordings \textit{Nat. Mach. Intell.} \textbf{5} 1097--1107. doi: 10.1038/s42256-023-00714-5.
\bibitem{moreira2025} Moreira J P C, Carvalho V R, Mendes E M A M, Fallah A, Sejnowski T J, Lainscsek C and Comstock L 2025 An open-access EEG dataset for speech decoding: exploring the role of articulation and coarticulation \textit{Sci. Data} \textbf{12} 1017. doi: 10.1038/s41597-025-05187-2.
\bibitem{dataset2025} Moreira J P C, Carvalho V R, Mendes E M A M, Fallah A, Sejnowski T J, Lainscsek C and Comstock L 2025 EEG dataset for speech decoding, version 1.0.1, OpenNeuro (dataset). doi: 10.18112/openneuro.ds006104.v1.0.1.
\bibitem{comstock2024} Comstock L, Carvalho V R, Lainscsek C, Fallah A and Sejnowski T J 2024 Transcranial magnetic stimulation facilitates neural speech decoding \textit{Brain Sci.} \textbf{14} 895. doi: 10.3390/brainsci14090895.
\bibitem{madishetty2026} Madishetty V 2026 CIPHER: Conformer-based Inference of Phonemes from High-density EEG, arXiv:2604.02362. doi: 10.48550/arXiv.2604.02362.
\bibitem{gramfort2014} Gramfort A \textit{et al.} 2014 MNE software for processing MEG and EEG data \textit{NeuroImage} \textbf{86} 446--460. doi: 10.1016/j.neuroimage.2013.10.027.
\bibitem{ledoit2004} Ledoit O and Wolf M 2004 A well-conditioned estimator for large-dimensional covariance matrices \textit{J. Multivar. Anal.} \textbf{88} 365--411. doi: 10.1016/S0047-259X(03)00096-4.
\bibitem{barachant2013} Barachant A, Bonnet S, Congedo M and Jutten C 2013 Classification of covariance matrices using a Riemannian-based kernel for BCI applications \textit{Neurocomputing} \textbf{112} 172--178. doi: 10.1016/j.neucom.2012.12.039.
\bibitem{congedo2017} Congedo M, Barachant A and Bhatia R 2017 Riemannian geometry for EEG-based brain--computer interfaces; a primer and a review \textit{Brain-Computer Interfaces} \textbf{4} 155--174. doi: 10.1080/2326263X.2017.1297192.
\bibitem{he2020} He H and Wu D 2020 Transfer learning for brain--computer interfaces: a Euclidean space data alignment approach \textit{IEEE Trans. Biomed. Eng.} \textbf{67} 399--410. doi: 10.1109/TBME.2019.2913914.
\bibitem{lawhern2018} Lawhern V J \textit{et al.} 2018 EEGNet: a compact convolutional neural network for EEG-based brain--computer interfaces \textit{J. Neural Eng.} \textbf{15} 056013. doi: 10.1088/1741-2552/aace8c.
\bibitem{schirrmeister2017} Schirrmeister R T \textit{et al.} 2017 Deep learning with convolutional neural networks for EEG decoding and visualization \textit{Hum. Brain Mapp.} \textbf{38} 5391--5420. doi: 10.1002/hbm.23730.
\bibitem{song2023} Song Y, Zheng Q, Liu B and Gao X 2023 EEG Conformer: convolutional transformer for EEG decoding and visualization \textit{IEEE Trans. Neural Syst. Rehabil. Eng.} \textbf{31} 710--719. doi: 10.1109/TNSRE.2022.3230250.
\bibitem{pernet2019} Pernet C R, Appelhoff S, Gorgolewski K J, Flandin G, Phillips C, Delorme A and Oostenveld R 2019 EEG-BIDS, an extension to the brain imaging data structure for electroencephalography \textit{Sci. Data} \textbf{6} 103. doi: 10.1038/s41597-019-0104-8.
\bibitem{markiewicz2021} Markiewicz C J \textit{et al.} 2021 The OpenNeuro resource for sharing of neuroscience data \textit{eLife} \textbf{10} e71774. doi: 10.7554/eLife.71774.
\bibitem{lotte2018} Lotte F, Bougrain L, Cichocki A, Clerc M, Congedo M, Rakotomamonjy A and Yger F 2018 A review of classification algorithms for EEG-based brain--computer interfaces: a 10 year update \textit{J. Neural Eng.} \textbf{15} 031005. doi: 10.1088/1741-2552/aab2f2.
\bibitem{roy2019} Roy Y, Banville H, Albuquerque I, Gramfort A, Falk T H and Faubert J 2019 Deep learning-based electroencephalography analysis: a systematic review \textit{J. Neural Eng.} \textbf{16} 051001. doi: 10.1088/1741-2552/ab260c.
\bibitem{varoquaux2017} Varoquaux G, Raamana P R, Engemann D A, Hoyos-Idrobo A, Schwartz Y and Thirion B 2017 Assessing and tuning brain decoders: cross-validation, caveats, and guidelines \textit{NeuroImage} \textbf{145} 166--179. doi: 10.1016/j.neuroimage.2016.10.038.
\bibitem{varoquaux2018} Varoquaux G 2018 Cross-validation failure: small sample sizes lead to large error bars \textit{NeuroImage} \textbf{180} 68--77. doi: 10.1016/j.neuroimage.2017.06.061.
\bibitem{kessler2025} Kessler R, Enge A and Skeide M A 2025 How EEG preprocessing shapes decoding performance \textit{Commun. Biol.} \textbf{8} 1039. doi: 10.1038/s42003-025-08464-3.
\bibitem{kriegeskorte2009} Kriegeskorte N, Simmons W K, Bellgowan P S F and Baker C I 2009 Circular analysis in systems neuroscience: the dangers of double dipping \textit{Nat. Neurosci.} \textbf{12} 535--540. doi: 10.1038/nn.2303.
\bibitem{varma2006} Varma S and Simon R 2006 Bias in error estimation when using cross-validation for model selection \textit{BMC Bioinformatics} \textbf{7} 91. doi: 10.1186/1471-2105-7-91.
\bibitem{nichols2017} Nichols T E \textit{et al.} 2017 Best practices in data analysis and sharing in neuroimaging using MRI \textit{Nat. Neurosci.} \textbf{20} 299--303. doi: 10.1038/nn.4500.
\bibitem{haufe2014} Haufe S, Meinecke F, G\"orgen K, D\"ahne S, Haynes J-D, Blankertz B and Bie{\ss}mann F 2014 On the interpretation of weight vectors of linear models in multivariate neuroimaging \textit{NeuroImage} \textbf{87} 96--110. doi: 10.1016/j.neuroimage.2013.10.067.
\bibitem{diliberto2015} Di Liberto G M, O'Sullivan J A and Lalor E C 2015 Low-frequency cortical entrainment to speech reflects phoneme-level processing \textit{Curr. Biol.} \textbf{25} 2457--2465. doi: 10.1016/j.cub.2015.08.030.
\bibitem{blankertz2011} Blankertz B, Lemm S, Treder M, Haufe S and M\"uller K-R 2011 Single-trial analysis and classification of ERP components---a tutorial \textit{NeuroImage} \textbf{56} 814--825. doi: 10.1016/j.neuroimage.2010.06.048.
\end{thebibliography}
\end{document}